# Drastic Minimization in the van der Waals Interaction with the Bottom Epitaxial Graphene Layer by the Diels-Alder Surface Chemistry of the Top Graphene Layer

Santanu Sarkar [$*]

Intel Corporation, Logic Technology Development Group, Ronler Acres Campus, Hillsboro, OR 97124, United States
*E-mail: santanu.sarkar@intel.com
[$]Former address: Center for Nanoscale Science and Engineering, Departments of Chemistry and Chemical & Environmental Engineering, University of California, Riverside, CA 92521, United States
*E-mail: ssark002@ucr.edu

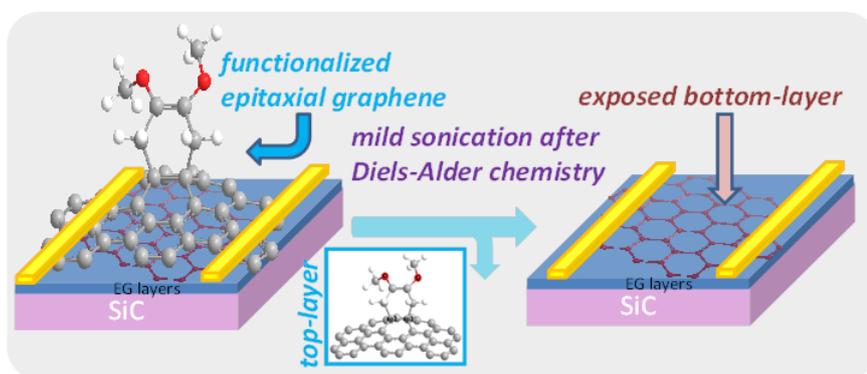

**ABSTRACT:** The Diels-Alder surface modified top epitaxial graphene layer, with newly created pair of $sp^3$ carbon centres, results in abrupt minimization of interlayer van der Waals interactions between two stacked graphene planes, and escapes the wafer during post-reaction manipulation stage, leaving the layer under it almost pristine-like. Above picture shows Diels-Alder functionalized $sp^2/sp^3$ graphene adduct leaves the parent wafer. In this communication we systematically address several fundamental questions in graphene surface chemistry, which are of extreme importance for device fabrication, and in successful implementation of covalently modified graphene in electronics industry.

Covalent chemistry of graphene at its unique Dirac point has great utility in engineering the electronic structures and magnetic properties of graphene.[1-4] A controlled chemical modification of the $sp^2$-honeycomb graphene network with a clear understanding of its edge structure is inevitable in implementing graphene into the post-silicon electronics technology.[5] The effect of new carbon-carbon bond formation chemistry and thereby subsequent generation of $sp^3$ carbon centers in graphene lattice is a well-established strategy in band-gap engineering of graphene.[6-10]

While covalent modification strategy is gaining increasing popularity from the standpoint of various applications (in band-gap engineering, edge smoothening, production of solution-processable graphene etc.)[6-12] a clear understanding of the effect of surface chemistry of graphene devices on the layer subjected to modification, and its neighbouring graphene layers is of fundamental value. Our present study (Scheme 1) demonstrates how the Diels-Alder surface chemistry of top epitaxial graphene layer (**1**) leads to significant reduction of the van der Waals interactions with the layer under it (**1'**); thereby the easing the removal of covalently modified top layer (**2**) leaving the bottom-layer pristine-like (unreacted, **1'**).

**Scheme 1.** (a,b) Schematic representation of the Diels-Alder adduct of top epitaxial graphene (EG) layer (**1**) with subsequent generation of a pair of A,B-$sp^3$ centers (**2**). (c) Schematics of the functionalized device, which easily loses the functionalized top layer (**2**), leaving the bottom layer pristine-like (unreacted, **1'**).

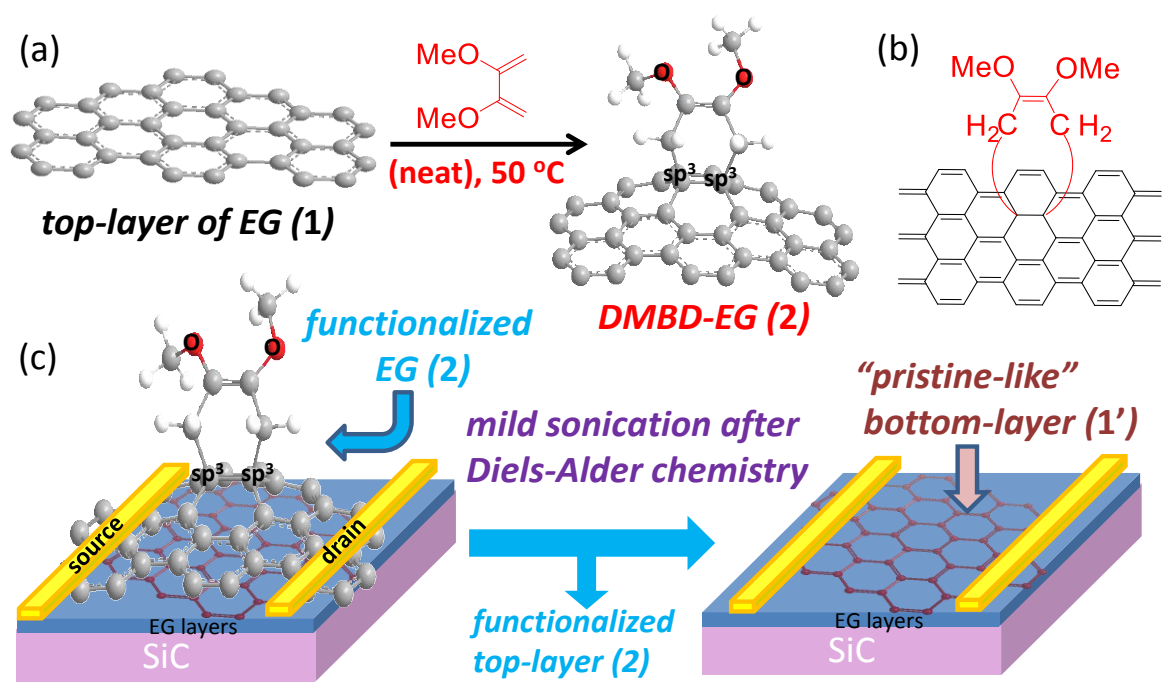

Wafer-scale growth of very large-area high-quality epitaxial graphene (EG)[13] wafers with typical wafer size of 4.5 × 3.5 mm$^2$ at Georgia Institute of technology (Prof. Walt de Heer Groups) has provided broad opportunities to explore unique reactivity of graphene as the solid state counterpart of classical small molecule chemistry, such as the Diels-Alder reaction.[3-4] Theory and experiment have shown that the planes in epitaxially-grown multilayer graphene are rotationally disordered and thus they are electronically decoupled and this preserves the electronic properties of an isolated graphene sheet in the vicinity of the Dirac point.[13] The versatility of graphene as a Diels-Alder substrate, behaving as both diene and dienophile is attributed to its unique zero-band-gap electronic structure[1] along with its possession of high-lying HOMO (low ionization potential; $E_{HOMO} = -IP$) and low-lying LUMO (high electron affinity; $E_{LUMO} = -EA$).[3-4] The work-function (W) of graphene ($E_{HOMO} = E_{LUMO} = -W$) is defined by the crossing of valence (HOMO) and conduction (LUMO) bands in graphene.[3]

Our present study has utilized the reactivity of high-quality classical epitaxial graphene (EG, C-face) surface as a dienophile and its reactions with a diene (here 2,3-dimethoxy-1,3-butadiene, DMBD) is employed as a model case-study to understand the effect of surface chemistry. This Diels-Alder chemistry leads to a pair of 1,2-$sp^3$ (non-planar tetrahedral) hybridized carbon centers in place of $sp^2$ (planar trigonal) hybridized carbon-based graphene lattice (necessarily at A and B sub-lattices) on the basal plane and edges of graphene.

In general, covalent modification of graphene via new carbon–carbon bond formation leads to an increase of the resistance of the sample.[5-6] In our prior measurements of temperature dependence of four-point resistances of pristine EG showed a slight decrease in resistance with temperature with a crossover to metallic behavior below 110 K (Figure 1a), while the covalently modified EG wafer (DMBD-EG) showed an activated temperature-dependence (Figure 1b) with non-metallic behavior over the whole temperature range (a characteristic of weak localization).[4] After cleaning the same DMBD-EG sample with solvents and mild sonication shows more than two fold decrease of its room-temperature resistance as compared to the pristine EG. This observation motivated us to pursue further studies to understand the effect of the Diels-Alder post-grafting manipulation. We pursued scanning tunnelling microscopy (STM) imaging, Raman spectroscopy and band mapping, atomic force microscopy (AFM) imaging to answer the fundamental questions about the wafer surface after solvent washings as follows: What happens to the $sp^3$-functioanlized top graphene layer? Does it escape the wafer during washings? If so, what happens to the graphene layer under it?

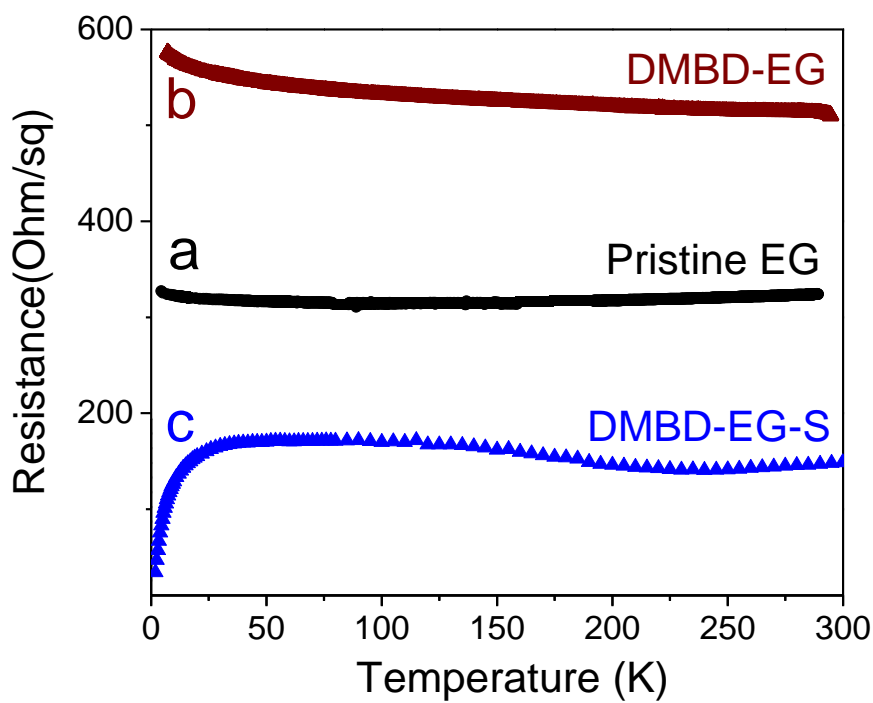

**Figure 1.** Temperature dependence of electrical transport (Ω vs T) of (a) pristine epitaxial graphene (EG-P), (b) DMBD-EG (after Diels-Alder chemistry), and (c) after solvent washing of DMBD-EG [DMBD-EG-S].

**In this communication we systematically explain our effort to address above fundamental questions, which are of extreme importance for device fabrication, and in successful implementation of covalently modified graphene in electronics industry.**

**Table 1. Data of temperature dependence of sheet resistances at various stages of epitaxial graphene wafer functionalization.**
[a] Pristine EG; [b] DMBD functionalized EG; [c] solvent-cleaned DMBD-EG.

| Temperature | EG-P[a] | DMBD-EG[b] | DMBD-EG-S[c] |
|---|---|---|---|
| 4 K | 328 Ω | 577 Ω | 121 Ω |
| 300 K | 324 Ω | 510 Ω | 145 Ω |

Scanning tunnelling microscopy (STM) and spectroscopy (STS) offers a direct probe to the electronic structure of pristine and chemically modified graphene.[3,14] The two-dimensional fast Fourier transform (2D-FFT) spectrum of a STM image of the solvent cleaned Diels-Alder modified EG (DMBD-EG-S) consists of the six outer bright spots from the graphene superlattice (Figure 2a, see the inset shown by a larger circle) and six spots corresponding to the graphene lattice in the center (Figure 2b, see the inset with a smaller circle).

Filtering higher order spots in the FFT spectrum (in Figure 2a) removes the noise in the STM images (which yields to Figure 2b), while filtering the graphene lattice (as in the inset of Figure 2b) removes everything inside of the largest circle circumscribed by the hexagon of the superlattice points yields Fig. 2c, which show the modified local denisty of states (LDOS). It can be seen in Figure 2d, that LDOS of pristine EG (EG-P, defect-free few-layer, ~3 layer EG)[14] and as-prepared DMBD-EG (Figure 2e) have striking differences,[3] while solvent-washed sample, DMBD-EG-S (Figure 2c, 2f) has essentially similar LDOS as that of pristine EG; only few point defects are observed in it.

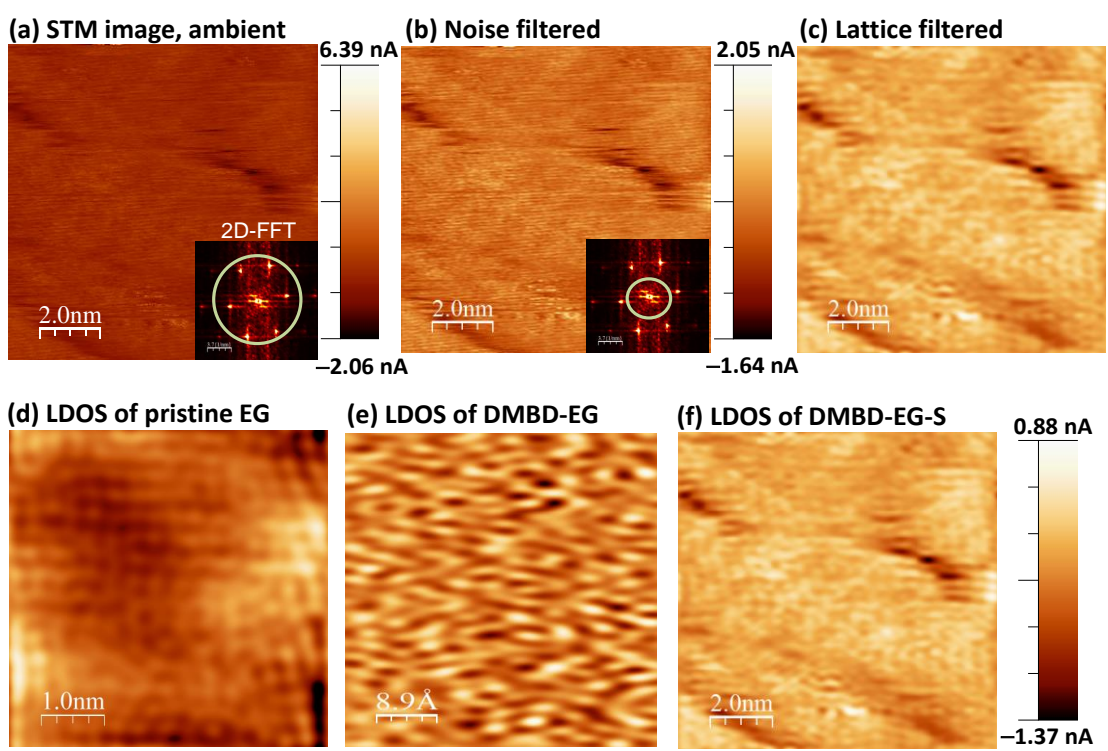

**Figure 2.** Scanning tunnelling microscopy (STM) images of solvent-cleaned DMBD-EG [DMBD-EG-S] under ambient conditions. Tunneling parameters: $V_{bias}$ = +5.0 mV, It = 3.9 nA. (a) Ambient STM image, (b) noise-filtered, and (c) lattice-filered STM iamge, which shows local density of states. Comparison of LDOSs of (a) pristine EG (EG-P), (b) DMBD-EG, and (c) of solvent-cleaned DMBD-EG.

Atomic force microscopy (AFM) images in the tapping (intermittent contact) mode of pristine EG, as-prepared DMBD-EG, and solvent cleaned sample (DMBD-EG-S) are compared in Figure 3. In can be estimated from AFM that the DMBD functional groups on flat EG surfaces has approximate height of 1.397 nm (Supporting Information), and surface functional of EG does not lead to the formation of any polymeic coating on EG; uniform functioanlization of both basal planes and edges of graphene are achieved (Figure 3b). Cleaning of DMBD-EG with solvent (mild-sonication) leads to removal of top functionalized graphene-layer, and shows almost clean, pristine-like (Figure 3a) graphene surface in DMBD-EG-S (Figure 3c).

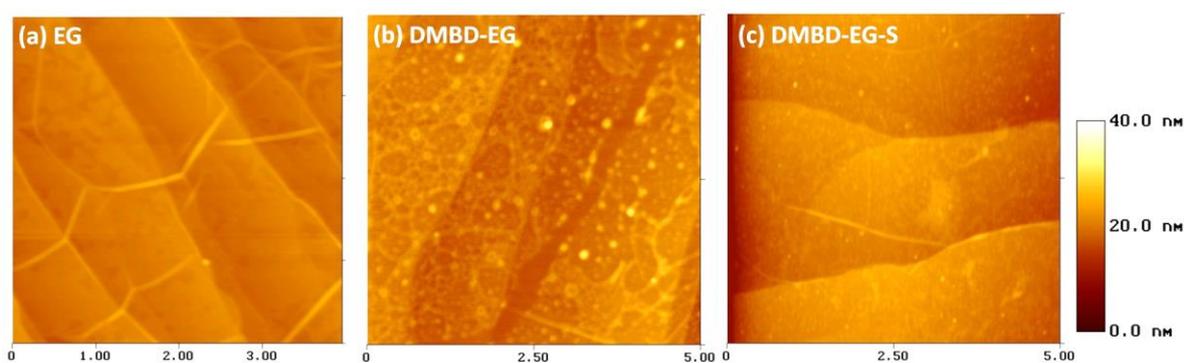

**Figure 3.** Comparison of atomic force microscopy (AFM) images of (a) pristine EG (EG-P), (b) DMBD-EG, and (c) of solvent-cleaned DMBD-EG [DMBD-EG-S]. The scale at the bottom of each figure is in μm.

Raman spectroscopy is an essential tool in graphene chemistry.[1-3] In Raman spectrscopy the relative ratios of integrated intensities of Raman D (at ~1345 cm$^{-1}$) to G (at ~1585 cm$^{-1}$) band ($I_D/I_G$) quantifies the relative content of the sp$^3$ carbon centers (presence of the D peak due to intervalley scattering) to the G band, and provides a useful index of the degree of chemical functionalization. The Diels-Alder adduct formation of EG leads to an increase of ($I_D/I_G$) ratio to 0.49 as compared to 0.03 in pristine EG. Washing DMBD-EG with solvent leads to the removal of functaionalized top layer, leaving bottom-layer almost defect-free (with $I_D/I_G$ = ~0.05). Additionally Raman D-band mapping of a selected area of graphene wafer provides useful information about the local degree of population (relative content) of sp$^3$ carbon centers (functional groups). Two dimensional Raman maps of the ratio of integrated area of D-band (centered at ~1355 cm$^{-1}$) to the integrated area of G-peak (centered at ~1581 cm$^{-1}$) with 1271 spectra each, at points spaced 1 μm apart, are collected in the selected 40 μm × 30 μm sample areas of the EG wafer for comparison and to obatin statistical information of functionalization homogeny. As can be seen in Figure 4b that DMBD-EG adduct has moderate to high degree of sp$^3$-centers over the whole wafer, while solvent washing the sample leads to removal of this functioanlzied top DMBD-EG layer, leaving almost pristine-like bottom layer (Figure 4c). The thereeree-dimensional (3D) Raman maps in Figure 4d show the spatial distribution of sp[3]-defects in the wafer; the bottom image shows only few defect sites which are almost negligible.

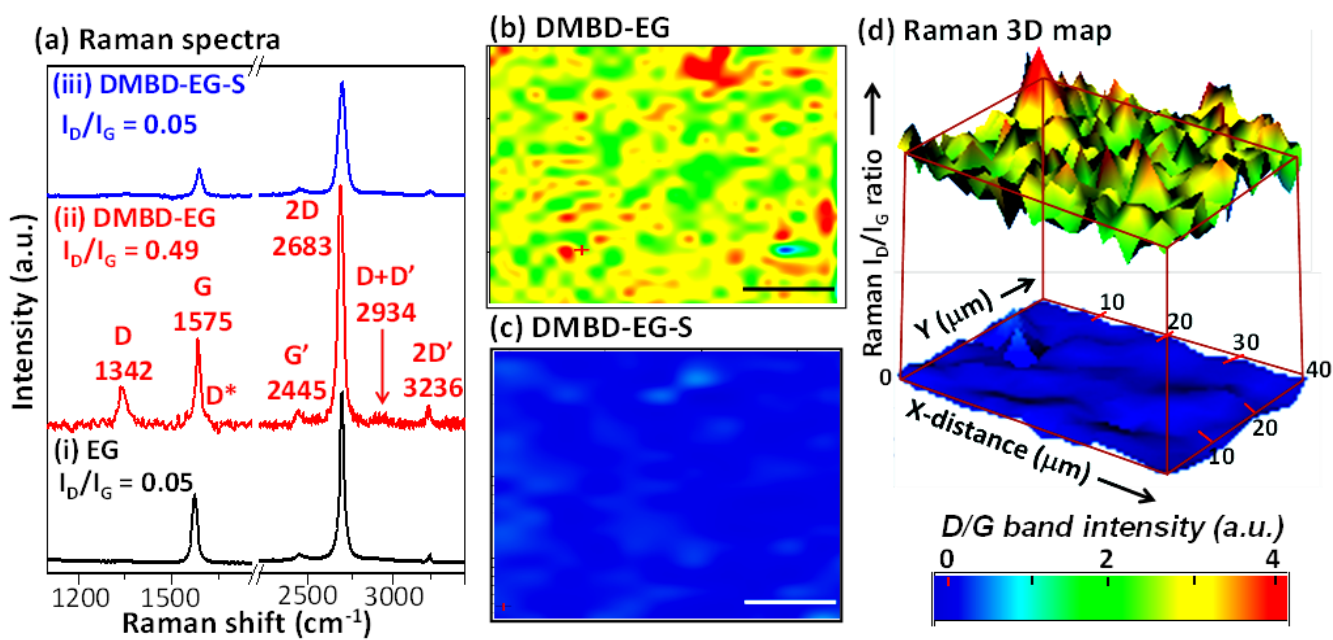

**Figure 4.** (a) Raman spectra (laser wavelength, $\lambda_{ex}$ = 532 nm, spot size = 0.7 μm) of (i) pristine EG (EG-P), (ii) DMBD-EG, and (iii) of solvent-cleaned DMBD-EG [DMBD-EG-S]. Two-dimensional Raman $I_D/I_G$ -band mapping of a selected 40 μm × 30 μm area of (b) DMBD-EG, showing moderate to high D-band intensity over the selected area, and (c) DMBD-EG-S, showing that removal of functionalized top layer leaves the graphene layer under it almost defect-free (pristine-like). Scale bar is 10 μm. (d) Raman 3D mapping of DMBD-EG (top) and the same sample after mild sonication, DMBD-EG-S (bottom).

In conclusion, we have demonstrated that the covalent Diels-Alder surface chemistry, which introduces a pair sp[3]-carbon center in the honeycomb lattice of graphene, significantly minimizes the van der Waals interactions between two adjacent graphene layers. This phenomenon is rationalized based on electrical transport measurement, STM imaging, AFM studies, and Raman spectroscopy and selected area Raman band mapping experiments. The unprecedented insights on the ability of functionalized top graphene layer to minimize its interaction with the layer under it and subsequent ease of being removed with gentle washing or mild sonication in an organic solvent will provide a cautionary note during handling and post-reaction processing stages in manipulating graphene devices and will aid to provide a rational basis for the design and implication of graphene in the post-CMOS manufacturing technology. The work will also provide a fundamental basis for chemical exfoliation of graphite to bulk functionalized graphene by covalent chemistry of the top graphite layer.[15]\

**Experimental Section**

**Materials:** Epitaxial graphene (EG) samples used for the present study are grown on single-crystal SiC(0001) on the C-face of a SiC wafer (Cree Inc, High purity R grade 4H SiC) by hydrogen etching to produce atomically flat surfaces and vacuum graphitization at ~1450 °C (UHV; base pressure $1 \times 10^{-10}$ Torr).

**Synthesis:** All surface chemical modifications and characterizations are performed on the C-face of EG. 2,3-dimethoxy-1,3-butadiene (DMBD, stored at 0-5 °C), p-xylene were all obtained from Sigma-Aldrich and reactions are performed under purified argon. In a typical reaction, the EG wafer (dienophile) is heated overnight at 50-55 °C in presence of ~0.1 M solution of DMBD (diene) in anhydrous p-xylene under argon, cleaned gently with neat p-xylene, then pentane and dried under slow flow of argon.[4] For solvent-cleaned samples (DMBD-EG-S), DMBD-EG (prepared as described before) is washed with acetone, then isopropanol, and again with acetone.

**Measurements and Characterizations:** The electrical transport and its temperature dependence data are acquired in a custom-made helium variable-temperature probe using a Lake Shore 340 temperature controlled driven with custom LabVIEW software. The device is degassed for 12 h in vacuum ($2 \times 10^{-8}$ Torr) prior to the measurement. The STM experiments are performed under ambient condition (room temperature, open atmosphere) and the images were acquired using a Digital Instruments Nanoscope IIIa multimode scanning probe microscope with commercial Pt/Ir tips. The processing (2D-FFT filtering) of the as-acquired STM images were performed using the WSxM software package.[Horcas, 2007 #5164] Raman spectra and the corresponding Raman mapping data were acquired using a Nicolet Almega XR Dispersive Raman microscope with a 0.7 micron spot size and a 532 nm laser excitation. The AFM images were acquired with the Digital Instruments Nanoscope IIIa instrument.

**References:**
(1) Sarkar, S.; Bekyarova, E.; Haddon, R. C. *Mater. Today* **2012,** 15, 276-285.
(2) Sarkar, S.; Bekyarova, E.; Haddon, R. C. *Angew. Chem. Int. Ed.* **2012,** 51, 4901-4904.
(3) Sarkar, S.; Bekyarova, E.; Haddon, R. C. *Acc. Chem. Res.* **2012,** 45, 673-682.
(4) Sarkar, S.; Bekyarova, E.; Niyogi, S.; Haddon, R. C. *J. Am. Chem. Soc.* **2011,** 133, 3324-3327.

(5) Bekyarova, E.; Sarkar, S.; Niyogi, S.; Itkis, M. E.; Haddon, R. C. *J. Phys. D: Appl. Phys.* **2012,** 45, 154009.

(6) Bekyarova, E.; Sarkar, S.; Wang, F.; Itkis, M. E.; Kalinina, I.; Tian, X.; Haddon, R. C. *Acc. Chem. Res.* **2013,** 46, 65-76.

(7) Wang, F.; Itkis, M. E.; Bekyarova, E.; Tian, X.; Sarkar, S.; Pekker, A.; Kalinina, I.; Moser, M.; Haddon, R. C. *Appl. Phys. Lett.* **2012,** 100, 223111.

(8) Wang, F.; Itkis, M. E.; Bekyarova, E.; Sarkar, S.; Tian, X.; Haddon, R. C. *J. Phys. Org. Chem.* **2012,** 25, 607-610.

(9) Kalinina, I.; Bekyarova, E.; Sarkar, S.; Wang, F.; Itkis, M. E.; Tian, X.; Niyogi, S.; Jha, N.; Haddon, R. C. *Macromol. Chem. Phys.* **2012,** 213, 1001-1019.

(10) Tian, X.; Sarkar, S.; Moser, M. L.; Wang, F.; Pekker, A.; Bekyarova, E.; Itkis, M. E.; Haddon, R. C. *Mater. Lett.* **2012,** 80, 171-174.

(11) (a) Sarkar, S.; Niyogi, S.; Bekyarova, E.; Haddon, R. C. *Chem. Sci.* **2011,** 2, 1326-1333.

(b) Haddon, R. C.; Sarkar, S.; Niyogi, S.; Bekyarova, E.; Itkis, M. E.; Tian, X.; Wang, F. Organometallic chemistry of extended periodic π-electron systems. U.S. Patent WO2012051597 A2, 13/879, 477, October 14, 2011.

(c) Sarkar, S. et al. Chem. Mater., 2014, 26 (1), pp 184–195 (DOI: 10.1021/cm4025809).

(12) Sarkar, S.; Zhang, H.; Huang, J.-W.; Wang, F.; Bekyarova, E.; Lau, C. N.; Haddon, R. C. Adv. Mater. 2013, 25, 1131-1136.

(13) (a) Berger, C.; Song, Z.; Li, T.; Li, X.; Ogbazghi, A. Y.; Feng, R.; Dai, Z.; Marchenkov, A. N.; Conrad, E. H.; First, P. N.; de Heer, W. A. J. Phys. Chem. B 2004, 108, 19912-19916.

(b) de Heer, W. A.; Berger, C.; Wu, X.; Sprinkle, M.; Hu, Y.; Ruan, M.; Stroscio, J.; First, P. N.; Haddon, R. C.; Piot, B.; Faugeras, C.; Potemski, M.; Moon, J.-S. J. Phys. D: Appl. Phys. 2010, 43, 374007.

(14) Niyogi, S.; Bekyarova, E.; Hong, J.; Khizroev, S.; Berger, C.; de Heer, W.; Haddon, R. C. J. Phys. Chem. Lett. 2011, 2 (19), 2487-2498.

(15) Englert, J. M. Et al. Nature Chem. 2011, 3, 279–286 (doi:10.1038/nchem.1010).

# Supporting Information

## Drastic Minimization in the van der Waals Interaction with the Bottom Epitaxial Graphene Layer by the Diels-Alder Surface Chemistry of the Top Graphene Layer

**Santanu Sarkar [$]\***

**Intel Corporation, Logic Technology Development, Ronler Acres Campus, Hillsboro, OR 97124, United States**
**\*E-mail:** santanu.sarkar@intel.com

[$]**Former address: Center for Nanoscale Science and Engineering, Departments of Chemistry and Chemical & Environmental Engineering, University of California, Riverside, CA 92521, United States**
**\*E-mail:** ssark002@ucr.edu

## Table of Contents

S1: Scanning Tunnelling Microscopy (STM)
S2: Additional AFM Images of Diels-Alder Functionalized Epitaxial Graphene
S3: Supporting References

Graphene has been shown to behave as both diene and dienophile in Diels-Alder chemistry due to its unique electronic structure [crossing of valence bands (HOMO) and conduction band (LUMO) at its Dirac point, resulting in zero-band-gap electronic structure] along with possession of exceptionally low ionization energy (high-lying HOMO) and very high electron affinity (low-lying LUMO).[3-4] In general, the Diels-Alder chemistry of graphene as a diene (its reaction with a dienophile, e.g. maleic anhydride) leads to 1,4-$sp^3$ carbon centers in graphene, while reactivity of graphene as a dienophile (its reaction with a diene, e.g. 2,3-dimethoxy-1,3-butadiene) leads to 1,2-$sp^3$ carbon centers in place of $sp^2$-honeycomb graphene carbon lattices. In the manuscript, we discuss the effect of the Diels-Alder surface chemistry of the exposed topmost layer epitaxial graphene (EG) layer on the layer under it, and subsequent ease of removal of the functionalized top-layer under mild sonication in solvents (isopropanol and acetone), resulting in exposure of the graphene underlayer almost pristine-like. This is demonstrated using STM, AFM, Raman spectroscopy and mapping, and electrical transport measurements (temperature-dependence of resistance).

**S1: Processing of STM Images of Pristine Epitaxial Graphene:**

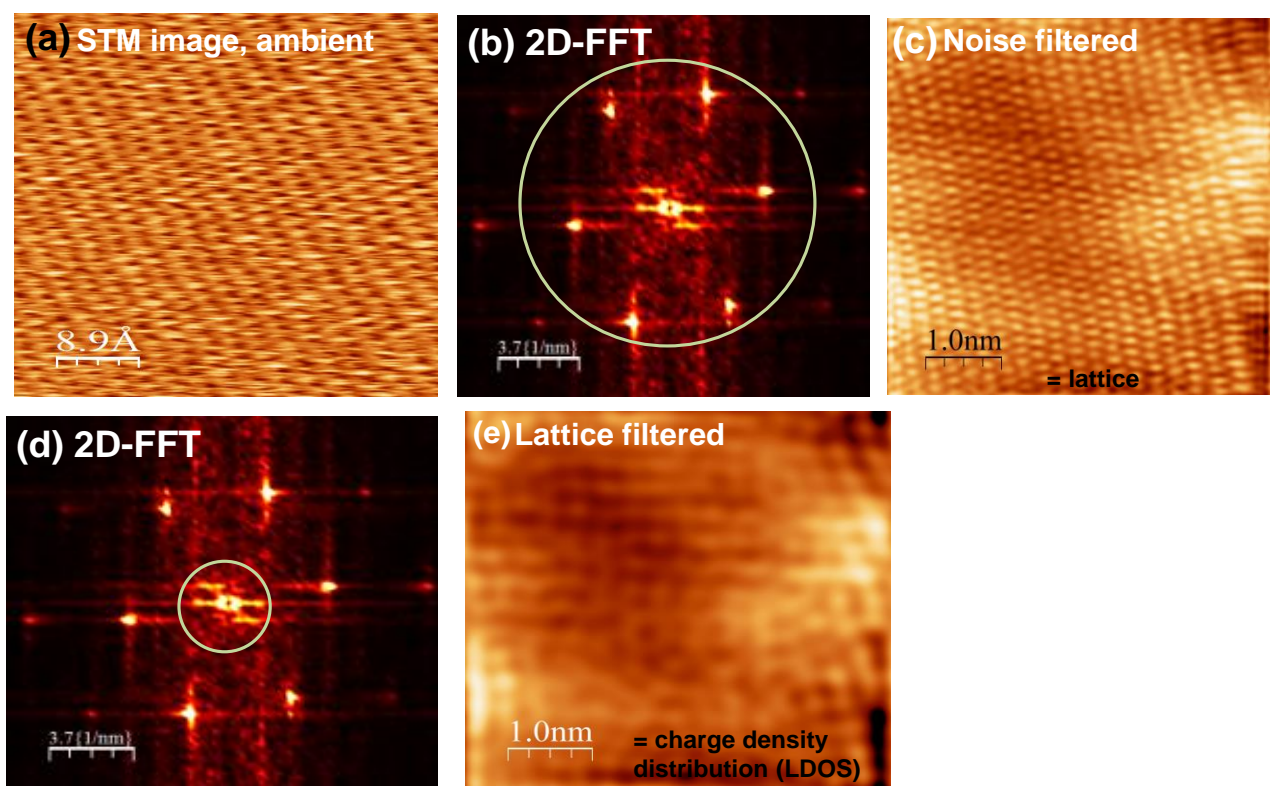

**Figure S1**. STM current images of pristine epitaxial graphene (EG) under ambient conditions. (a) Pristine EG, $V_{bias}$ = +5.1 mV, $I_t$ = 4.3 nA. (b) 2D-FFT spectrum of the STM image. (c) STM image of EG after subtracting noise. (d) 2D-FFT filtering of lattice and (e) the resulting STM image of EG after lattice filtering showing the LDOS.

**S2: Additional AFM Images of Diels-Alder Functionalized Epitaxial Graphene:** Tapping mode AFM imaging were acquired to microscopically visualize the uniformity of surface coverage (functionalization) and to analyze the height of the DMBD functional groups in the resulting Diels-Alder adduct with epitaxial graphene (EG). The height of the attached functional groups is approximately estimated as 1.397 nm and no polymeric coating on the EG surface was found (**Figure S2**). Step edges of grown EG are obvious from **Figure S3**, where the differences in vertical distances between one graphene layer to the next graphene layer varies by 9.5557 nm.

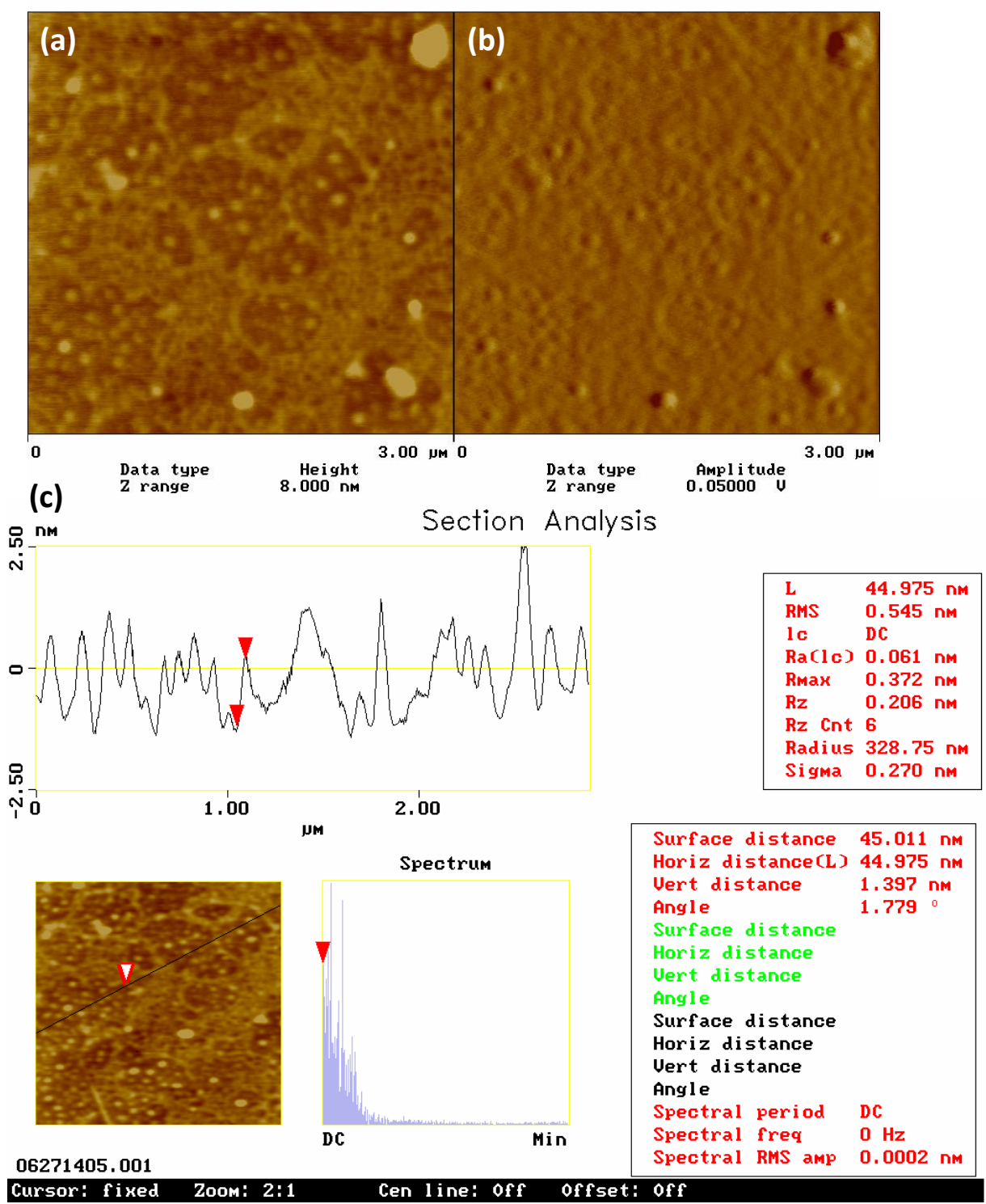

**Figure S2**. Atomic force microscopy (AFM) images of DMBD functioanlized EG, showing the height of the attached functional groups is approximately estimated as 1.397 nm and no polymeric coating on the EG surface.

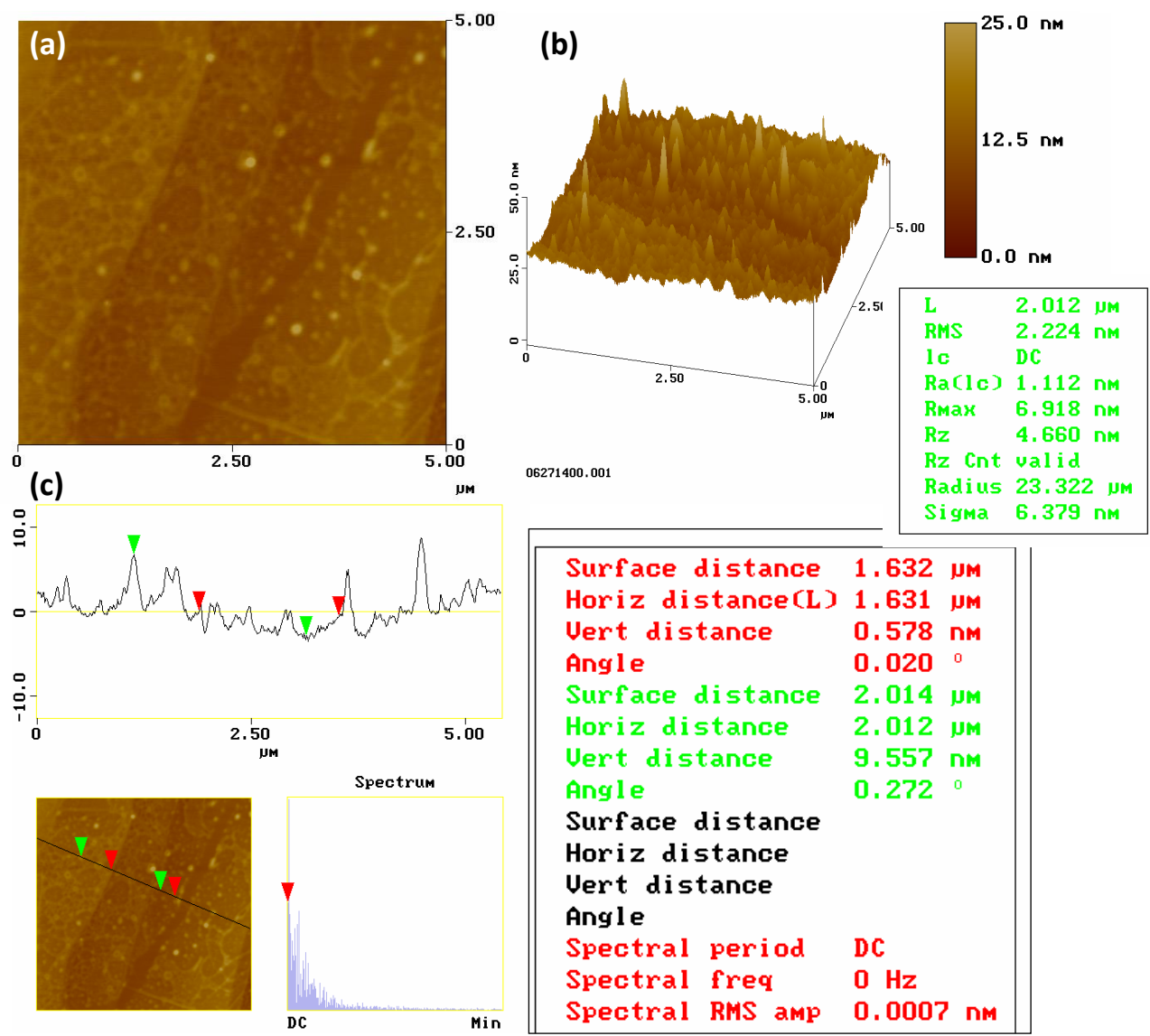

**Figure S3**. AFM images of DMBD functioanalied EG samples, in which where the differences in vertical distances between one graphene layer to the next graphene layer varies by 9.5557 nm.

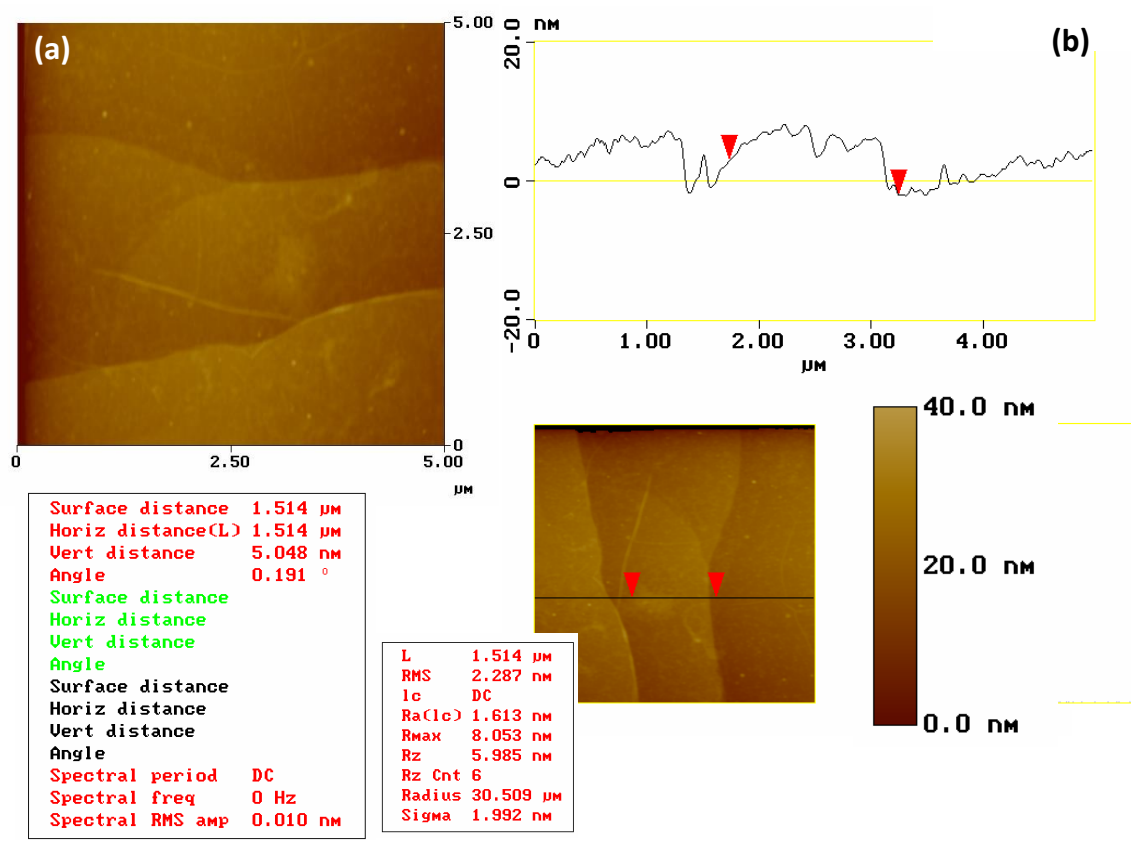

**Figure S4.** AFM iamges of pristine EG showing step edges of grown EG.

## Supporting References:

(1) Sarkar, S.; Bekyarova, E.; Haddon, R. C. *Mater. Today* **2012,** 15, 276-285.

(2) Sarkar, S.; Bekyarova, E.; Haddon, R. C. *Angew. Chem. Int. Ed.* **2012,** 51, 4901-4904.

(3) Sarkar, S.; Bekyarova, E.; Haddon, R. C. *Acc. Chem. Res.* **2012,** 45, 673-682.

(4) Sarkar, S.; Bekyarova, E.; Niyogi, S.; Haddon, R. C. *J. Am. Chem. Soc.* **2011,** 133, 3324-3327.

(5) Bekyarova, E.; Sarkar, S.; Niyogi, S.; Itkis, M. E.; Haddon, R. C. *J. Phys. D: Appl. Phys.* **2012,** 45, 154009.

(6) Bekyarova, E.; Sarkar, S.; Wang, F.; Itkis, M. E.; Kalinina, I.; Tian, X.; Haddon, R. C. *Acc. Chem. Res.* **2013,** 46, 65-76.

(7) Wang, F.; Itkis, M. E.; Bekyarova, E.; Tian, X.; Sarkar, S.; Pekker, A.; Kalinina, I.; Moser, M.; Haddon, R. C. *Appl. Phys. Lett.* **2012,** 100, 223111.

(8) Wang, F.; Itkis, M. E.; Bekyarova, E.; Sarkar, S.; Tian, X.; Haddon, R. C. *J. Phys. Org. Chem.* **2012,** 25, 607-610.

(9) Kalinina, I.; Bekyarova, E.; Sarkar, S.; Wang, F.; Itkis, M. E.; Tian, X.; Niyogi, S.; Jha, N.; Haddon, R. C. *Macromol. Chem. Phys.* **2012,** 213, 1001-1019.

(10) Tian, X.; Sarkar, S.; Moser, M. L.; Wang, F.; Pekker, A.; Bekyarova, E.; Itkis, M. E.; Haddon, R. C. *Mater. Lett.* **2012,** 80, 171-174.

(11) (a) Sarkar, S.; Niyogi, S.; Bekyarova, E.; Haddon, R. C. *Chem. Sci.* **2011,** 2, 1326-1333.

    (b) Haddon, R. C.; Sarkar, S.; Niyogi, S.; Bekyarova, E.; Itkis, M. E.; Tian, X.; Wang, F. Organometallic chemistry of extended periodic π-electron systems. U.S. Patent WO2012051597 A2, 13/879, 477, October 14, 2011.

    (c) Sarkar, S. et al. Chem. Mater., 2014, 26 (1), pp 184–195 (DOI: 10.1021/cm4025809).

(12) Sarkar, S.; Zhang, H.; Huang, J.-W.; Wang, F.; Bekyarova, E.; Lau, C. N.;  Haddon, R. C. Adv. Mater. 2013, 25, 1131-1136.

(13) (a) Berger, C.; Song, Z.; Li, T.; Li, X.; Ogbazghi, A. Y.; Feng, R.; Dai, Z.; Marchenkov, A. N.; Conrad, E. H.; First, P. N.; de Heer, W. A. J. Phys. Chem. B 2004, 108, 19912-19916.

    (b) de Heer, W. A.; Berger, C.; Wu, X.; Sprinkle, M.; Hu, Y.; Ruan, M.; Stroscio, J.; First, P. N.; Haddon, R. C.; Piot, B.; Faugeras, C.; Potemski, M.; Moon, J.-S. J. Phys. D: Appl. Phys. 2010, 43, 374007.

(14) Niyogi, S.; Bekyarova, E.; Hong, J.; Khizroev, S.; Berger, C.; de Heer, W.; Haddon, R. C. J. Phys. Chem. Lett. 2011, 2 (19), 2487-2498.

(15) Englert, J. M. Et al.  Nature Chem. 2011, 3, 279–286 (doi:10.1038/nchem.1010).